\documentclass[article,aps,pra,reprint,twocolumn,superscriptaddress,10pt]{revtex4-2}
\usepackage{amsmath}%,preprint
\usepackage{mathrsfs}
\usepackage{amsfonts}
\usepackage{amssymb}
\usepackage{graphicx}
\usepackage[colorlinks,linkcolor=blue,anchorcolor=blue,urlcolor=blue,citecolor=blue]{hyperref}
\usepackage{eufrak}
\usepackage{multirow}
\usepackage{mathrsfs}
\usepackage{textcomp}
\usepackage{appendix}
\usepackage{makecell}
\usepackage{multirow}
\usepackage{xcolor}

\begin{document}

\title{Linear optical fan-out gates using fewer ancillary single photons with enhanced success probability}

\author{Wen-Qiang Liu}\email{Contact author: wqliu@stdu.edu.cn}
\affiliation{Department of Mathematics and Physics, Shijiazhuang Tiedao University, Shijiazhuang 050043, China}
\author{Hai-Rui Wei}\email{Contact author: hrwei@ustb.edu.cn}
\affiliation{School of Mathematics and Physics, University of Science and Technology Beijing, Beijing 100083, China}

\date{\today }

\begin{abstract}
Photonic quantum gates are fundamental building blocks for a wide range of optical quantum information processing tasks. We propose an efficient linear-optical scheme for implementing a post-selected three-qubit fan-out gate (also known as a controlled-NOT-NOT gate) using only two ancillary single photons and linear optical elements. The scheme can be generalized to an $n$-qubit fan-out gate, requiring $(n-1)$ ancillary single photons and $(3n-3)$ polarizing beam splitters (PBSs), with a success probability of $\left(\frac{1}{4}\right)^{n-1}$. Compared to the standard gate decomposition approach, which requires $(2n-2)$ ancillary single  photons, $(5n-5)$ PBSs, and yields a success probability of $\left(\frac{1}{8}\right)^{n-1}$. Our scheme significantly reduces the resource overhead and improves the success probability. We further evaluate the gate performance and demonstrate improved robustness compared to gate decomposition-based methods.
\end{abstract}

%\pacs{03.67.Lx, 42.50.Ex, 05.70.Ln} and

\maketitle

\section{Introduction}  \label{sec1}

Photonic quantum computing  \cite{kok2007linear}, which utilizes photons as information carriers for quantum state encoding, manipulation, and storage \cite{o2007optical}, has emerged as a leading platform for demonstrating quantum computational advantage. Quantum computing  with photon qubits has shown remarkable superiority over classical systems in boson sampling experiments \cite{zhong2020quantum,deng2023gaussian,liu2024efficiently} and random circuit sampling tasks \cite{gao2025establishing}. The inherent advantages of photons, including long coherence times, multiple accessible degrees of freedom, and high-fidelity single-qubit operations \cite{psiquantum2025manufacturable}, make them particularly suitable for various quantum computing applications \cite{couteau2023applications,ding2025high,main2025distributed}.
However, in photonic quantum information processing, a fundamental challenge persists as the implementation of deterministic multi-photon quantum gates remains elusive due to the weak nonlinear interactions between individual photons \cite{uppu2021quantum}. This limitation results in inherently probabilistic gate operations, significantly hindering the scalability of optical quantum computing systems. Consequently, developing resource-efficient methods to enhance the success probability of multi-photon quantum gates has become a critical research focus in the noisy intermediate-scale quantum (NISQ) era \cite{memon2024quantum}.

The realization of two-qubit entangling gates using linear optics and photon detection was first demonstrated in the KLM scheme \cite{knill2001scheme}, which has since inspired extensive efforts to implement controlled-NOT (CNOT) gates or their equivalent controlled-phase flip (CPF) gates \cite{he2023super}. Nowadays, experimental implementations of CNOT gates in multiphoton systems can generally be divided into three categories. In the first category, no ancillary photons are required, and a success probability of 1/9 of post-selected CNOT gate  \cite{ralph2002linear}  based on linear optics has been reported \cite{o2003demonstration,langford2005demonstration,kiesel2005linear,lemr2011experimental,he2013demand}. The second category improves the success probability to 1/4 by employing an additional pair of maximally entangled photons \cite{knill2001scheme}, and this approach represents the highest success probability to date \cite{pittman2001probabilistic,gasparoni2004realization,zhao2005experimental,zhou2011adding,zeuner2018integrated}. Considering the challenges and imperfections associated with generating entangled photon pairs \cite{orieux2017semiconductor}, a third category of schemes based on ancillary single photons has been theoretically proposed and experimentally demonstrated. In this category, CPF gates with success probabilities of 1/16 \cite{knill2001scheme} and
2/27 \cite{knill2002quantum} first were  proposed. By  introducing two ancillary single photons and Bell-state measurements, the success probability of the CNOT gate was further improved to
1/8 \cite{bao2007optical,li2021heralded}. Beyond these basic gates, research has also extended to more complex multiple-photon gates like Toffoli and Fredkin gates, though these typically exhibit lower success probabilities \cite{fiuravsek2006linear,gong2008methods,lanyon2009simplifying,kieling2010photonic,liu2020low,li2022quantum,liu2022universal,li2022chip,liu2023linear}.

Fan-out gates \cite{baumer2025measurement}, commonly known as multiple-target CNOT gates, represent a fundamental and powerful class of multiqubit controlled operations \cite{hoyer2005quantum}. An $n$-qubit fan-out gate, denoted as controlled-NOT-$\cdots$-NOT or $C^n$ gate, operates on one control qubit and ($n-1$) target qubits. These gates serve as essential components in quantum information processing with critical applications including quantum error correction \cite{nielsen2010quantum}, fault-tolerant quantum computing \cite{postler2022demonstration}, entangled state generation \cite{yang2020implementing,liu2025deterministic}, measurement-based quantum computing \cite{raussendorf2001one}, and circuit optimization \cite{gokhale2021quantum}. Recent experimental advances have successfully realized fan-out gates in both Rydberg atom systems \cite{khazali2020fast,young2021asymmetric,wei2022fast,stolz2022quantum} and photonic systems \cite{xu2023multiqubit}. For photonic implementations, current schemes rely heavily on multiphoton entangled Greenberger-Horne-Zeilinger (GHZ) states as ancillary resources \cite{xu2023multiqubit}. However, generating large-scale photonic GHZ states with high efficiency remains a significant experimental challenge. Although an $n$-qubit fan-out gate can theoretically be decomposed into ($n-1$) CNOT gates sharing one control qubit \cite{baumer2025measurement}, this approach becomes impractical for large-scale optical implementations due to its substantial resource requirements and heightened sensitivity to noise.

In this paper, we first propose a post-selected three-photon fan-out $C^3$ gate assisted by two ancillary single photons by using only linear optical elements. This scheme is then generalized to realize an arbitrary $n$-qubit fan-out $C^n$ gate with the assistance of ($n-1$) ancillary single photons and ($3n-3$) polarizing beam splitters (PBSs). The gate operation succeeds when each output path contains exactly one photon, yielding an overall success probability of $\left(\frac{1}{4}\right)^{n-1}$. Notably, the previous scheme \cite{xu2023multiqubit} requires an ancillary maximally entangled GHZ state, whereas our approach eliminates this requirement and relies only on ancillary single photons. We further analyze the impact of imperfect PBSs on the fidelity of the $C^n$ gate and demonstrate that our scheme exhibits greater robustness to such imperfections and achieves a higher success probability compared to the gate decomposition-based method \cite{baumer2025measurement}.

\section{multiphoton fan-out gates with linear optics} \label{Sec2}

\subsection{Post-selected three-photon $C^3$ gate}  \label{Sec2.1}

The three-qubit $C^3$ gate with one control qubit and two target qubits,  is the simplest fan-out quantum gate. The operator of  $C^3$ gate is given by
\begin{eqnarray}
\begin{split}             \label{eq1}
&U_{C^3}= \left(\begin{array}{ccccc}
             I_4  &   0 &   0 &   0 &   0  \\
             0    &   0 &   0 &   0 &   1  \\
             0    &   0 &   0 &   1 &   0  \\
             0    &   0 &   1 &   0 &   0  \\
             0    &   1 &   0 &   0 &   0  \\
\end{array}\right).\\
\end{split}
\end{eqnarray}
Here $I_{4}$ denotes the four-order identity operator. The $C^3$ gate describes if and only if the control qubit is in the state $|1\rangle$, two target qubit will qubit-flip and the other case, the two target qubits remain unchanged.

\begin{figure} %[!h]   %[htpb]
\begin{center}
\includegraphics[width=7.3 cm,angle=0]{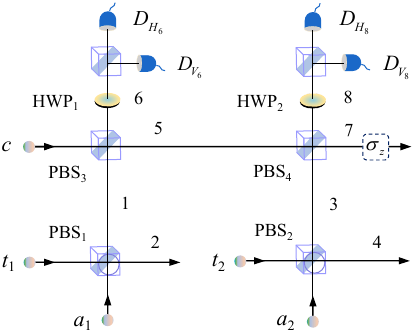}
\caption{Schematic diagram for implementing a post-selected photonic $C^3$ gate in output paths 7, 2, and 4. PBS$_1$ and PBS$_2$ are diagonal polarizing beam splitters in the basis $\{|\pm\rangle = \frac{1}{\sqrt{2}}(|H\rangle \pm |V\rangle)\}$, while PBS$_3$ and PBS$_4$ are the regular polarizing beam splitters in the basis  $\{|H\rangle, |V\rangle\}$. HWP$_1$ and HWP$_2$ are the half-wave plates set at 22.5$^\circ$. The diagonal PBS can be realized by inserting HWPs set at 22.5$^\circ$ at two inputs and two outputs of the regular PBS, respectively. The numbered labels denote the optical paths of the photons.} \label{figure1}
\end{center}
\end{figure}

Figure \ref{figure1} schematically illustrates the implementation of a three-photon $C^3$ gate using two ancillary single photons, several PBSs, and the post-selection technique.
The computational basis in our scheme is encoded in the horizontal polarization state $|H\rangle$ and the vertical polarization state $|V\rangle$.
Let us explain the scheme in details below.

As shown in Fig.~\ref{figure1}, the initial control qubit and the two target qubits are prepared in general single-photon polarization states, which are given by
\begin{eqnarray}             \label{eq2}
|\psi\rangle_c = \alpha_1|H\rangle_{c}+\beta_1|V\rangle_{c},
\end{eqnarray}
\begin{eqnarray}             \label{eq3}
|\psi\rangle_{t_1} = \alpha_2|H\rangle_{t_1}+\beta_2|V\rangle_{t_1},
\end{eqnarray}
\begin{eqnarray}             \label{eq4}
|\psi\rangle_{t_2} = \alpha_3|H\rangle_{t_2}+\beta_3|V\rangle_{t_2}.
\end{eqnarray}
Here the complex coefficients satisfy $|\alpha_i|^2+|\beta_i|^2=1$ for $i$=1, 2, 3. Two ancillary single photons are prepared in the horizontal polarization state
\begin{eqnarray}             \label{eq5}
|\psi\rangle_{a_1} =|H\rangle_{a_1},
\end{eqnarray}
\begin{eqnarray}             \label{eq6}
|\psi\rangle_{a_2} =|H\rangle_{a_2}.
\end{eqnarray}
Therefore, the input five single-photon states in the whole optical system can be expressed as
\begin{eqnarray}
\begin{split}             \label{eq7}
|\Psi\rangle_\text{in} =|\psi\rangle_{c}\otimes|\psi\rangle_{t_1}\otimes|\psi\rangle_{a_1}\otimes|\psi\rangle_{t_2}\otimes|\psi\rangle_{a_2}.
\end{split}
\end{eqnarray}

First, the photon in path $t_1$ and the first ancillary photon in path $a_1$ simultaneously undergo a diagonal PBS$_1$, which operates in the diagonal polarization basis $\{|\pm\rangle=\frac{1}{\sqrt{2}}(|H\rangle\pm|V\rangle)\}$, transmitting the $|+\rangle$ component and reflecting the $|-\rangle$ component.  Similarly, the photon in path $t_2$ and the second ancillary photon in path $a_2$ pass through a PBS$_2$ with the same diagonal polarization operation. After the photons from paths $t_1$ and $a_1$ arrive simultaneously at PBS$_1$ and those from paths $t_2$ and $a_2$ arrive simultaneously at PBS$_2$, the initial state $|\Psi\rangle_\text{in}$ becomes
\begin{eqnarray}
\begin{split}             \label{eq8}
|\Psi\rangle_1 = &\; \frac{1}{4} (\alpha_1|H\rangle_{c}+\beta_1|V\rangle_{c})\otimes
[\alpha_2(|+\rangle_2+|-\rangle_1) \\& +\beta_2(|+\rangle_2-|-\rangle_1)]\otimes(|+\rangle_1+|-\rangle_2)\\&
\otimes[\alpha_3(|+\rangle_4+|-\rangle_3)  +\beta_3(|+\rangle_4-|-\rangle_3)]\\&
\otimes(|+\rangle_3+|-\rangle_4).
\end{split}
\end{eqnarray}
To improve readability, Eq. \eqref{eq8} can be rewritten in a more compact form as
\begin{eqnarray}
\begin{split}             \label{eq9}
|\Psi\rangle_1 =&\; \frac{1}{4} (\alpha_1|1_H\rangle_{c}+\beta_1|1_V\rangle_{c})
\otimes\big\{\alpha_2\big[|1_H\rangle_1|1_H\rangle_2 \\& +|1_V\rangle_1|1_V\rangle_2
+\frac{1}{2}(|2_H\rangle_1-|2_V\rangle_1+|2_H\rangle_2\\& -|2_V\rangle_2)\big]
% %
+\beta_2\big[|1_H\rangle_1|1_V\rangle_2 +|1_V\rangle_1|1_H\rangle_2\\&
+\frac{1}{2}(|2_H\rangle_2-|2_V\rangle_2-|2_H\rangle_1
+|2_V\rangle_1)\big]\big\}\\&
\otimes\big\{\alpha_3\big[|1_H\rangle_3|1_H\rangle_4+|1_V\rangle_3|1_V\rangle_4\\&
+\frac{1}{2}(|2_H\rangle_3
-|2_V\rangle_3
+|2_H\rangle_4 -|2_V\rangle_4)\big]\\&
% %
+\beta_3\big[|1_H\rangle_3|1_V\rangle_4
+|1_V\rangle_3|1_H\rangle_4\\&
+\frac{1}{2}(|2_H\rangle_4-|2_V\rangle_4 -|2_H\rangle_3 +|2_V\rangle_3)\big]\big\}.
\end{split}
\end{eqnarray}
Here the notation $|n_H\rangle_k$ ($|n_V\rangle_k$) denotes a photonic state containing $n$ $H$-polarized ($V$-polarized) photons in path $k$. For simplicity, we drop the photon number index when describing states post-selected for single photons. When exactly one photon exist in each of paths 1, 2, 3, and 4 after PBS$_1$ and PBS$_2$, Eq.~\eqref{eq9} reduces to the following normalized state
\begin{eqnarray}
\begin{split}             \label{eq10}
|\Psi\rangle_2 = &\; \frac{1}{2} (\alpha_1|H\rangle_{c}+\beta_1|V\rangle_{c})\otimes
[\alpha_2(|H\rangle_1|H\rangle_2 \\& +|V\rangle_1|V\rangle_2)
+\beta_2(|H\rangle_1|V\rangle_2 +|V\rangle_1|H\rangle_2)]\\&
\otimes [\alpha_3(|H\rangle_3|H\rangle_4 +|V\rangle_3|V\rangle_4) \\&
+\beta_3(|H\rangle_3|V\rangle_4  +|V\rangle_3|H\rangle_4)]
\end{split}
\end{eqnarray}
with a probability of $\frac{1}{4}$.

Second, the photons in paths $c$ and 1 are interfered at regular PBS$_3$, which is aligned in the $\{|H\rangle, |V\rangle\}$ basis and transmits $H$-polarized photons while reflecting $V$-polarized ones. The PBS$_3$ can post-select the even-party states $|H\rangle_5|H\rangle_6$ and $|V\rangle_5|V\rangle_6$ when exactly one photon is present in each of paths 5 and 6.
Subsequently, the photon in path 5 and the photon in path 3 are interfered at PBS$_4$. Similarly, the PBS$_4$ can also post-select the even-party states $|H\rangle_7|H\rangle_8$ and $|V\rangle_7|V\rangle_8$ when only one photon in paths 7 and 8. Therefore, when exactly one photon exists in paths 5, 6, 7, and 8 after the PBS$_3$ and PBS$_4$, the state of the whole system is collapsed into
\begin{eqnarray}
\begin{split}             \label{eq11}
|\Psi\rangle_3 = &\; \alpha_1\alpha_2\alpha_3|H\rangle_6|H\rangle_8|H\rangle_7|H\rangle_2|H\rangle_4 \\&
+\alpha_1\alpha_2\beta_3|H\rangle_6|H\rangle_8|H\rangle_7|H\rangle_2|V\rangle_4 \\&
+\alpha_1\beta_2\alpha_3|H\rangle_6|H\rangle_8|H\rangle_7|V\rangle_2|H\rangle_4 \\&
+\alpha_1\beta_2\beta_3|H\rangle_6|H\rangle_8|H\rangle_7|V\rangle_2|V\rangle_4 \\&
+\beta_1\alpha_2\alpha_3|V\rangle_6|V\rangle_8|V\rangle_7|V\rangle_2|V\rangle_4 \\&
+\beta_1\alpha_2\beta_3|V\rangle_6|V\rangle_8|V\rangle_7|V\rangle_2|H\rangle_4 \\&
+\beta_1\beta_2\alpha_3|V\rangle_6|V\rangle_8|V\rangle_7|H\rangle_2|V\rangle_4 \\&
+\beta_1\beta_2\beta_3|V\rangle_6|V\rangle_8|V\rangle_7|H\rangle_2|H\rangle_4
\end{split}
\end{eqnarray}
with a probability of $\frac{1}{4}$.

Finally, the photons in paths 6 and 8 pass through half-wave plates (HWP$_1$ and HWP$_2$) set at $22.5^\circ$, which each implements a Hadamard operation in the $\{|H\rangle, |V\rangle\}$ basis
\begin{eqnarray}
\begin{split}             \label{eq12}
&|H\rangle \xrightarrow{\text{HWP}} \frac{1}{\sqrt{2}} (|H\rangle+|V\rangle), \\
&|V\rangle \xrightarrow{\text{HWP}} \frac{1}{\sqrt{2}} (|H\rangle-|V\rangle).
\end{split}
\end{eqnarray}
After the actions of HWP$_1$ and HWP$_2$, the five-photon state $|\Psi\rangle_3$ evolves into
\begin{eqnarray}
\begin{split}             \label{eq13}
|\Psi\rangle_4 = &\; \frac{1}{2}\big[I_7C^3(|\psi\rangle_7|\psi\rangle_2|\psi\rangle_4)\otimes |H\rangle_6|H\rangle_8 \\&
+\sigma_{z7}C^3(|\psi\rangle_7|\psi\rangle_2|\psi\rangle_4)\otimes |H\rangle_6|V\rangle_8\\&
+\sigma_{z7}C^3(|\psi\rangle_7|\psi\rangle_2|\psi\rangle_4)\otimes |V\rangle_6|H\rangle_8\\&
+I_7C^3(|\psi\rangle_7|\psi\rangle_2|\psi\rangle_4)\otimes |V\rangle_6|V\rangle_8 \big].
\end{split}
\end{eqnarray}
Here the product state $|\psi\rangle_7|\psi\rangle_2|\psi\rangle_4=(\alpha_1|H\rangle_{7}+\beta_1|V\rangle_{7})\otimes(\alpha_2|H\rangle_{2}+\beta_2|V\rangle_{2})\otimes(\alpha_3|H\rangle_{4}+\beta_3|V\rangle_{4})$ represents the output state in paths 7, 2, and 4, identical to the initial input state in paths $c$, $t_1$, and $t_2$. $I_7$ is the identity operation acting on the photon in path 7, and $\sigma_{z7}$ is Pauli $Z$ operation acting on the photon in path 7.

\begin{table} [tb]
\renewcommand{\arraystretch}{1.4}  % 增大行间距
\centering \caption{The correspondence between the measurement outcomes of the ancillary photons in paths 6 and 8 and the required feed-forward operations on the output photons in paths 7, 2, and 4 for implementing the $C^3$ gate. The symbol “\textbf{--}” indicates that no feed-forward operation is required.}
\begin{tabular}{cccccccccccccc}%{lp{1.2in} lp{1.2in}}
\hline\hline

\multicolumn {2}{c}{Measurement}  & \;\;\;  & \multicolumn {3}{c}{\;\; Feed-forward}     \\ \cline{1-2}  \cline{4-6}

Path 6   &  \;    Path 8         &\;\;\;   & \;  Path 7    &\qquad \;\; Path 2     & \qquad \;\; Path 4                    \\

\hline

$D_{H_6}$  & \; $D_{H_8}$    &\;\;\;   & \;  --   &\qquad \;\;  --     & \qquad \;\; --                    \\

$D_{V_6}$  & \; $D_{V_8}$    &\;\;\;   & \;  --   &\qquad \;\;  --     & \qquad \;\;  --                  \\

$D_{H_6}$  & \; $D_{V_8}$    &\;\;\;   & \; $\sigma_z$    &\qquad \;\;  --   & \qquad \;\; --                    \\

$D_{V_6}$  & \; $D_{H_8}$     &\;\;\;   & \; $\sigma_z$   &\qquad \;\;  --   & \qquad \;\; --                  \\
\hline \hline
\end{tabular}\label{Table1}
\end{table}

Based on Eq. \eqref{eq13}, a two-photon coincidence measurement is performed on the photons in paths 6 and 8. The measurement outcomes and the corresponding feed-forward operations are presented in Tab. \ref{Table1}. If the coincidence detectors $D_{H_6}$--$D_{H_8}$ or $D_{V_6}$--$D_{V_8}$ click, then a three-photon $C^3$ gate is implemented on the output photons in paths 7, 2, and 4, where the photon in path 7 serves as the control qubit and those in paths 2 and 4 serve as the target qubits. Alternatively, if the coincidence detectors $D_{H_6}$--$D_{V_8}$ or $D_{V_6}$--$D_{H_8}$ click, the $C^3$ gate is still obtained by applying a feed-forward Pauli-$Z$ operation to the photon in path 7. The gate operation succeeds if and only if exactly one photon is present in each output path (2, 4, 6, 7, and 8); otherwise, the scheme fails. Notably, the success of the scheme cannot be heralded solely by the responses of the detectors setting in paths 6 and 8. But the instances in which only the path 6 or 8 contains one photon or multiple photons or no photons indicate that the scheme fails. The total success probability of the scheme is therefore given by $\frac{1}{4} \times \frac{1}{4} = \frac{1}{16}$.

\subsection{Post-selected $n$-photon $C^n$ gate}  \label{Sec2.2}

\begin{figure} [tpb] %[!h]   %
\begin{center}
\includegraphics[width=8.2 cm,angle=0]{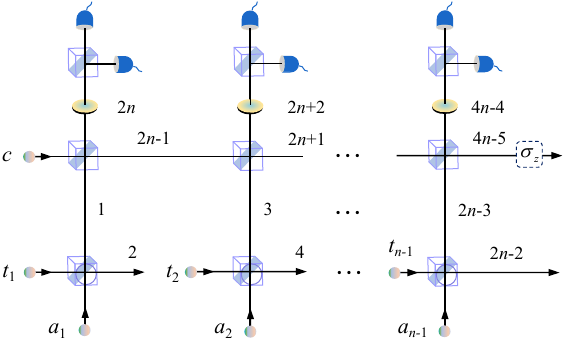}
\caption{Schematic diagram for implementing an $n$-photon $C^n$ gate in the paths $4n-5, 2, 4, \ldots, 2n-2$. Here the photon in path $4n-5$ serves as the control qubit, while the photons in paths $2, 4, \ldots, 2n-2$ serve as the $n-1$ target qubits.}
\label{figure2}
\end{center}
\end{figure}

Our method can be scaled to implement an $n$-photon fan-out $C^n$ gate. Figure~\ref{figure2} schematically illustrates the implementation of such a gate using $(n-1)$ ancillary single photons and post-selection techniques. The $n$ input photons, located in paths $c$, $t_1$, $\ldots$, $t_{n-1}$, are prepared in a general product state given by
\begin{eqnarray}
\begin{split}             \label{eq14}
|\psi\rangle_{c, t_1, \ldots, t_{n-1}} = &\; (\alpha_1|H\rangle_{c}+\beta_1|V\rangle_{c}) \\&
\bigotimes_{j=1}^{n-1}
(\alpha_{j+1}|H\rangle_{t_j}+\beta_{j+1}|V\rangle_{t_j}).
\end{split}
\end{eqnarray}
Here the complex coefficients satisfy the normalization condition $|\alpha_i|^2+|\beta_i|^2=1$ for $i=1, 2, \ldots, n$. Each of the $(n-1)$ ancillary photons is prepared in the horizontal polarization state $|H\rangle$, i.e.,
\begin{eqnarray}             \label{eq15}
|\psi\rangle_{a_j} =|H\rangle_{a_j}.
\end{eqnarray}
Therefore, the joint initial state of the entire system is expressed as
\begin{eqnarray}
\begin{split}             \label{eq16}
|\Psi'\rangle_\text{in} =&\;  (\alpha_1|H\rangle_{c}+\beta_1|V\rangle_{c})\\&
\bigotimes_{j=1}^{n-1}
(\alpha_{j+1}|H\rangle_{t_j}+\beta_{j+1}|V\rangle_{t_j})|H\rangle_{a_j}.
\end{split}
\end{eqnarray}

\textbf{Step 1:} Similar to the argument in Sec.~\ref{Sec2.1}, each of the $n-1$ ancillary photons interferes with one of the $n-1$ target photons at a diagonal PBS in the diagonal basis $\{|\pm\rangle\}$. Each PBS performs a parity-check operation, it routes photons with the same polarization into different output paths and those with orthogonal polarizations into the same output path. After the $n-1$ PBSs, when exactly one photon appears in each output path, the system evolves into the normalized state
\begin{eqnarray}
\begin{split}             \label{eq17}
|\Psi'\rangle_1 =&\; 2^{-\frac{n-1}{2}} (\alpha_1|H\rangle_{c}+\beta_1|V\rangle_{c})\\&
\bigotimes_{j=1}^{n-1}
\bigg[\alpha_{j+1}(|H\rangle_{2j-1}|H\rangle_{2j}+|V\rangle_{2j-1}|V\rangle_{2j})\\&
+\beta_{j+1}(|H\rangle_{2j-1}|V\rangle_{2j}+|V\rangle_{2j-1}|H\rangle_{2j})
\bigg]
\end{split}
\end{eqnarray}
with a success probablitity of $\frac{1}{2^{n-1}}$.

\textbf{Step 2:} The control photon in path $c$ sequentially interacts with the photons in paths $1, 3, \dots, 2n-3$ via a series of $n-1$ PBSs  in the $\{|H\rangle, |V\rangle\}$ basis. For each PBS, only the even-parity components $|H\rangle|H\rangle$ and $|V\rangle|V\rangle$ are preserved when exactly one photon exits each output port. Under this condition, the state in Eq.~\eqref{eq17} reduces into
\begin{eqnarray}
\begin{split}             \label{eq18}
|\Psi'\rangle_2 =&\; \alpha_1|H\rangle_{2n-5}\bigotimes_{j=1}^{n-1}
\bigg[\alpha_{j+1}(|H\rangle_{2j}+|V\rangle_{2j})\\& \otimes |H\rangle_{2(n+j-1)}\bigg]
+\beta_1|V\rangle_{2n-5} \\& \bigotimes_{j=1}^{n-1}\bigg[\alpha_{j+1}(|V\rangle_{2j}+|H\rangle_{2j})\otimes |V\rangle_{2(n+j-1)}\bigg]
\end{split}
\end{eqnarray}
with a success probability of $\frac{1}{2^{n-1}}$.

\textbf{Step 3:} All photons in paths $2n$, $2n+2$, $\ldots$, $4n-4$, pass through the HWPs set at 22.5$^\circ$. These HWPs can realize the Hadamard operations on photons $H$ and $V$, and evolve the state in Eq. \eqref{eq18} into
\begin{eqnarray}
\begin{split}             \label{eq19}
|\Psi'\rangle_3 =&\; 2^{-\frac{n-1}{2}}\bigg[\sum_{\vec{\mu} \in \mathcal{E}} \big(|\vec{\mu}\rangle \otimes C^n
|\psi\rangle_{4n-5, 2, \ldots, 2n-2}\big)   \\&
+\sum_{\vec{\mu} \in \mathcal{O}} \big(|\vec{\mu}\rangle \otimes \sigma_{z 4n-5} C^n
|\psi\rangle_{4n-5, 2, \ldots, 2n-2}\big)  \bigg].
\end{split}
\end{eqnarray}
Here $\vec{\mu} = (\mu_{2n}, \mu_{2n+2}, \dots, \mu_{4n-4})$ is an $(n-1)$-dimensional bit string with each $\mu_j \in \{H, V\}$. The sets $\mathcal{E}$ and $\mathcal{O}$ respectively contain strings with an even and odd number of $|V\rangle$ components. $|\psi\rangle_{4n-5, 2, \ldots, 2n-2}$ is a general $n$-photon state in the output ports $4n-5, 2, \ldots, 2n-2$, consistent with the initial input in Eq.~\eqref{eq14}.

\begin{figure} [tpb] %[!h]   %
\begin{center}
\includegraphics[width=5.3 cm,angle=0]{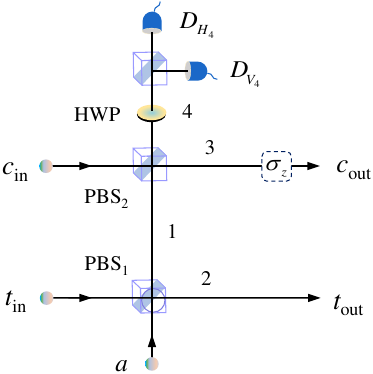}
\caption{Schematic diagram for implementing a two-photon $C^n$ gate with $n=2$ (i.e., CNOT gate).}
\label{figure3}
\end{center}
\end{figure}

Based on Eq.~\eqref{eq19}, the $C^n$ gate can be implemented in the output paths $4n-5, 2, \ldots, 2n-2$ by measuring the ancillary photons in paths $2n, 2n+2, \ldots, 4n-4$. If the number of $V$-photon in the measurement outcome $|\vec{\mu}\rangle$ is even, then no feed-forward operation is required. Conversely, if the number of $V$-photon outcomes is odd, then a Pauli-$Z$ operation will be applied to the photon in path $4n-5$ (the control qubit). The $C^n$ gate succeeds only when exactly one photon emitting from each output path ($4n-5, 2, \ldots, 2n-2, 2n, \ldots, 4n-4$). Otherwise, the gate operation fails. The success of the scheme cannot be heralded solely by detecting photons in the paths ($2n, 2n+2, \ldots, 4n-4$). The total success probability of the post-selected scheme is therefore given by $\frac{1}{2^{n-1}}\times\frac{1}{2^{n-1}}=\frac{1}{4^{n-1}}$.
For $n=2$, the fan-out $C^n$ gate simplifies to the linear optical CNOT gate with a success probability of $\frac{1}{4}$, see Fig.~\ref{figure3}.

\section{Performance of the scheme} \label{Sec3}

Our scheme primarily relies on a series of PBSs. Under ideal conditions, the scheme achieves unit fidelity. However, in realistic implementations, imperfections in PBSs inevitably degrade the gate performance. In practice, PBSs can be functionally replaced by calcite beam displacers (BDs), as both devices spatially separate photons based on their polarization. According to the experimentally characterized imperfections of BDs \cite{meng2022deterministic,jiang2024efficient}, we analyze how deviations from ideal PBS behavior affect the overall fidelity of the proposed scheme.

The imperfections of a PBS for polarization-encoded photons can be characterized by its non-ideal behavior in different polarization bases. For an imperfect PBS with a finite extinction ratio and angular deviation, its action can be modeled as the following unitary transformations

\begin{eqnarray}
\begin{split}             \label{eq20}
|H\rangle \xrightarrow{\hat{U}_\text{PBS}} &\frac{1}{\sqrt{1+|r|}} \big[(\cos\theta-\sqrt{r}\sin\theta)|H\rangle \\&-(\sin\theta+\sqrt{r}^*\cos\theta)|V\rangle\big],
\end{split}
\end{eqnarray}
\begin{eqnarray}
\begin{split}             \label{eq21}
|V\rangle \xrightarrow{\hat{U}_\text{PBS}} &\frac{1}{\sqrt{1+|r|}}  \big[(\sin\theta+\sqrt{r}\cos\theta)|H\rangle \\&+(\cos\theta-\sqrt{r}^*\sin\theta)|V\rangle\big],
\end{split}
\end{eqnarray}
\begin{eqnarray}
\begin{split}             \label{eq22}
|+\rangle \xrightarrow{\hat{U}_\text{PBS}} &\frac{1}{\sqrt{1+|r|}} \big[(\cos\theta-\sqrt{r}\sin\theta)|+\rangle \\&-(\sin\theta+\sqrt{r}^*\cos\theta)|-\rangle\big],
\end{split}
\end{eqnarray}
\begin{eqnarray}
\begin{split}             \label{eq23}
|-\rangle \xrightarrow{\hat{U}_\text{PBS}} &\frac{1}{\sqrt{1+|r|}} \big[(\sin\theta+\sqrt{r}\cos\theta)|+\rangle \\&+(\cos\theta-\sqrt{r}^*\sin\theta)|-\rangle\big].
\end{split}
\end{eqnarray}
Here the $\theta$ denotes the deviation of mirror mounts and $r$ denotes the polarization extinction ratio of the PBS. The term $\sqrt{r}^*$ is the complex conjugate of $\sqrt{r}$. Under these imperfect conditions, suppose the initial two-photon state
\begin{eqnarray}
\begin{split}             \label{eq24}
|\Psi\rangle=(\alpha_1|H\rangle_c+\beta_1|V\rangle_c)\otimes(\alpha_2|H\rangle_1+\beta_2|V\rangle_1)
\end{split}
\end{eqnarray}
undergoes an imperfect PBS$_3$ as shown in Fig.~\ref{figure1}, the real output state is given by
\begin{eqnarray}
\begin{split}             \label{eq25}
|\Psi_\text{real}\rangle=&\;\frac{1}{1+|r|}\big[\alpha_1\alpha_2\big((\cos\theta-\sqrt{r}\sin\theta)|H\rangle_5 %
\\&\;-(\sin\theta+\sqrt{r}^*\cos\theta)|V\rangle_5\big)
%%%%
\\&\;\times\big((\cos\theta-\sqrt{r}\sin\theta)|H\rangle_6 %
\\&\;-(\sin\theta+\sqrt{r}^*\cos\theta)|V\rangle_6\big)
%%%%
\\&\;+\alpha_1\beta_2\big((\cos\theta-\sqrt{r}\sin\theta)|H\rangle_5
\\&\;-(\sin\theta+\sqrt{r}^*\cos\theta)|V\rangle_5\big)
\\&\;\times\big((\sin\theta+\sqrt{r}\cos\theta)|H\rangle_5
\\&\;+(\cos\theta-\sqrt{r}^*\sin\theta)|V\rangle_5\big)
%%%%
\\&\;+\beta_1\alpha_2\big((\sin\theta+\sqrt{r}\cos\theta)|H\rangle_6
\\&\;+(\cos\theta-\sqrt{r}^*\sin\theta)|V\rangle_6\big)
\\&\;\times\big((\cos\theta-\sqrt{r}\sin\theta)|H\rangle_6
\\&\;-(\sin\theta+\sqrt{r}^*\cos\theta)|V\rangle_6\big)
\\&\;+\beta_1\beta_2\big((\sin\theta+\sqrt{r}\cos\theta)|H\rangle_6 %
\\&\;+(\cos\theta-\sqrt{r}^*\sin\theta)|V\rangle_6\big)
%%%%
\\&\;\times\big((\sin\theta+\sqrt{r}\cos\theta)|H\rangle_5 %
\\&\;+(\cos\theta-\sqrt{r}^*\sin\theta)|V\rangle_5\big)
\big].
\end{split}
\end{eqnarray}
While the ideal output state after the PBS$_3$ is
\begin{eqnarray}
\begin{split}             \label{eq26}
|\Psi_\text{ideal}\rangle=(\alpha_1|H\rangle_5+\beta_1|V\rangle_6)\otimes(\alpha_2|H\rangle_6+\beta_2|V\rangle_5).
\end{split}
\end{eqnarray}

\begin{figure} [tpb] %[!h]   %
\begin{center}
\includegraphics[width=8.2 cm,angle=0]{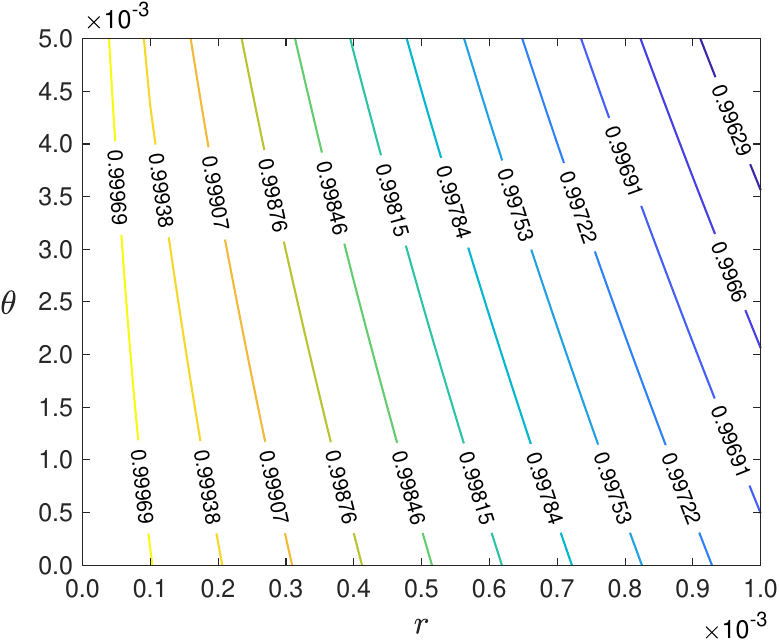}
\caption{The average fidelity of an imperfect PBS as a function of the angular deviation $\theta$ and the polarization extinction ratio $r$.}
\label{figure4}
\end{center}
\end{figure}

To quantify the performance of the $C^n$ gate affected by the imperfect PBSs, we first calculate the average fidelity of a PBS, which is given by
\begin{eqnarray}
\begin{split}             \label{eq27}
\overline{F}_\text{PBS}& =\frac{1}{(2\pi)^2}\int_{0}^{2\pi}\int_{0}^{2\pi}|\langle\Psi_\text{real}|\Psi_\text{ideal}\rangle|^2dxdy \\&
=\frac{1}{64 (1 + r)^2}[-20 (r^2-1) \cos(2\theta) \\ &\quad +  19 (r^2-6r +1) \cos(4\theta) + 5 (1 + r) \\ &\quad \times(5 + 5r - 8 \sqrt{r} \sin(2\theta)) \\ &\quad + 76 (r-1) \sqrt{r} \sin(4\theta)].
\end{split}
\end{eqnarray}
Here we set $\alpha_1=\cos x$, $\beta_1=\sin x$, $\alpha_2=\cos y$, and $\beta_2=\cos y$.
Figure~\ref{figure4} plots the average fidelity $\overline{F}_\text{PBS}$ as a function of the imperfection parameters $\theta$ and $r$. As expected, the fidelity improves as both $\theta$ and $r$ decrease. Specifically, for $\theta = 2 \times 10^{-3}$~rad and $r = 0.2 \times 10^{-3}$, the average fidelity reaches $\overline{F}_\text{PBS} = 0.9992$.
In our implementation of the $C^3$ gate shown in Fig.~\ref{figure1}, a total of six PBSs are required, operating in either the $\{|H\rangle, |V\rangle\}$ or $\{|+\rangle, |-\rangle\}$ basis. Assuming all PBSs exhibit the same performance, the overall average fidelity of the gate is estimated as $\overline{F}_{C^3}=(0.9992)^6=0.9952$.

\begin{table*}[tb]
\renewcommand{\arraystretch}{1.4}  % 增大行间距
\centering
\caption{Comparison between our scheme and the decomposition-based $C^n$ gate ($n \geq 3$), in terms of the required number of PBSs, the number of ancillary single photons (ASP), the number of ancillary entangled photon pairs (AEPP), the average fidelity $\overline{F}_{C^{\it{n}}}$, the success probability (SP), and nondestructiveness.}\label{Table2}
\begin{tabular}{lccccccc}
\hline\hline
Schemes &\;\; PBS &\;\; ASP &\;\;AEPP &\;\; $\overline{F}_{C^{\it{n}}}$ &\;\; SP  &\;\; nondestructive\\
\hline
GDM with \cite{langford2005demonstration,kiesel2005linear}  &\;\; $5n - 5$   &\;\; 0 &\;\; 0 &\;\; $0.9992^{5n - 5}$ &\;\; $\left(\frac{1}{9}\right)^{n - 1}$ &\;\; $\times$\\
GDM with \cite{bao2007optical,li2021heralded}   &\;\; $5n - 5$   &\;\; $2(n - 1)$ &\;\; 0 &\;\; $0.9992^{5n - 5}$ &\;\; $\left(\frac{1}{8}\right)^{n - 1}$ &\;\; \checkmark\\
GDM with \cite{pittman2001probabilistic,gasparoni2004realization,zhao2005experimental}   &\;\; $4n - 4$   &\;\; 0 &\;\; $n - 1$ &\;\; $0.9992^{4n - 4}$ &\;\; $\left(\frac{1}{4}\right)^{n - 1}$ &\;\; \checkmark\\
Ours  &\;\; $3n - 3$   &\;\; $n - 1$  &\;\; 0 &\;\; $0.9992^{3n - 3}$ &\;\; $\left(\frac{1}{4}\right)^{n - 1}$ &\;\; $\times$\\
\hline\hline
\end{tabular}
\end{table*}

\begin{figure} [t] %[!h]   %
\begin{center}
\includegraphics[width=8.2 cm,angle=0]{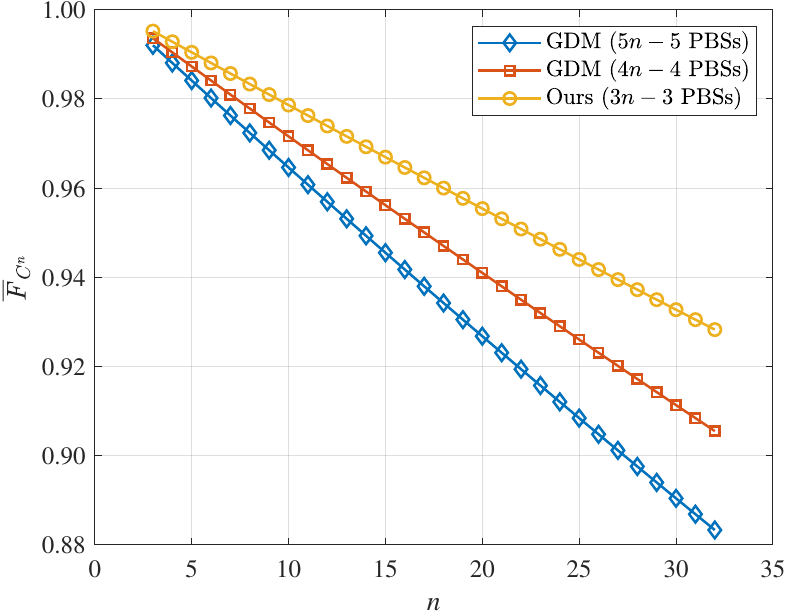}
\caption{Comparison of the average fidelities of the $C^n$ ($n\geq3$) gate between our scheme and the conventional gate decomposition method (GDM).}
\label{figure5}
\end{center}
\end{figure}

The $n$-qubit $C^n$ gate ($n \geq 3$) shown in Fig. \ref{figure2} requires $(3n - 3)$ PBSs and $(n - 1)$ ancillary photons. The average fidelity of our scheme is $0.9992^{3n - 3}$ and the corresponding success probability is $\left(\frac{1}{4}\right)^{n - 1}$.  In contrast, the gate decomposition method (GDM) \cite{baumer2025measurement} requires $(n - 1)$ CNOT gates to construct an $n$-qubit $C^n$ gate. Previous schemes for implementing optical CNOT gates can be categorized into three groups: (i) without ancillary photons, using five PBSs, and the success probability is $\frac{1}{9}$ \cite{langford2005demonstration,kiesel2005linear}; (ii) using an entangled photon pair as an ancillary resource, requiring four PBSs and achieving a success probability of $\frac{1}{4}$ \cite{pittman2001probabilistic,gasparoni2004realization,zhao2005experimental}; and (iii) using five PBSs and two ancillary single photons, yielding a success probability of $\frac{1}{8}$ \cite{bao2007optical,li2021heralded}. A detailed comparison between our scheme and the GDM-based $C^n$ gate is presented in Tab.~\ref{Table2}, and Fig.~\ref{figure5} illustrates the corresponding average fidelities of the $C^n$ gate for these approaches. Recently, a long-distance photonic $C^n$ gate assisted by an $n$-photon GHZ state has been proposed \cite{xu2023multiqubit}. In contrast, our scheme does not require a shared GHZ state as an ancillary resource, which significantly relaxes the experimental requirements.

Besides imperfect PBSs, other experimental imperfection sources that may degrade the performance of the schemes mainly include finite detector efficiency  (since implementing a
$C^n$ gate requires $2(n-1)$ detectors), imperfections in single-photon sources, detector dark counts, and nonideal behaviors of other optical elements.

\section{Conclusion}  \label{sec4}

We have presented a linear-optical scheme for implementing a three-photon $C^3$ gate using only two ancillary photons. This scheme is further generalized to an $n$-qubit fan-out $C^n$ gate with $(n-1)$ ancillary single photons and $(3n - 3)$ PBSs. The gate operates successfully when exactly one photon in each output path, yielding a total success probability of $\left(\frac{1}{4}\right)^{n - 1}$. The success of the gate cannot be completely heralded by detecting only the ancillary photons.  Instead, the success can be inferred either from subsequent practical applications or from measurements on all output photons. One possible way to overcome the destructive measurement of post-selection is to develop nondestructive photonic qubit detection \cite{o2016nondestructive,niemietz2021nondestructive} or to employ more ancillary photons (see Tab. \ref{Table2}). Compared with the GDM-based schemes and existing photonic gates, our scheme requires fewer ancillary single photons and fewer PBSs, while achieving higher fidelity and success probability.
Our results provide a more efficient and scalable alternative for realizing multi-qubit fan-out gates in linear-optical quantum computing, with promising applications in photonic quantum information processing.

\section*{ACKNOWLEDGMENTS}

This work was supported by the National Natural Science Foundation of China under Grant No. 12505028 and Grant No. 62371038, and  Science Research Project of Hebei Education Department under Grant No. QN2025054.

%\bibliography{mybibliography}


\end{document}